\documentclass[preprint,epsf,12pt]{aastex}
\pdfoutput=1
\begin{document}
\shorttitle{The star ingesting luminosity of IMBHs}

\shortauthors{Ramirez-Ruiz \& Rosswog}

\title{The star ingesting luminosity of intermediate mass black holes
  in globular clusters}

\author{Enrico Ramirez-Ruiz\altaffilmark{1} and Stephan
  Rosswog\altaffilmark{2}} \altaffiltext{1}{Department of Astronomy
  and Astrophysics, University of California, Santa Cruz, CA 95064}
\altaffiltext{2}{School of Engineering and Science, Jacobs University
  Bremen, Campus Ring 1, 28759 Bremen, Germany}

\begin{abstract} 
The dynamics of stars in the inner regions of nearby globular clusters
(GCs) such as G1 indicate the presence of central concentrated dark
masses, and one would like to know whether these are indeed
intermediate mass black holes (IMBHs). As the number of surrounding
stars, and their motions, are roughly known, the capture rate can be
estimated; the question then arises of whether the apparent quiescence
of the nuclei of these GCs is compatible with the IMBH's presence. The
role of debris from disrupted stars in {\it activating} quiescent
nuclei of GCs is studied here employing three-dimensional
hydrodynamics simulations. It is argued that when individual stars are
disrupted, the bulk of the debris would be swallowed or expelled
rapidly compared with the interval between successive disruptions.  A
transient (predominantly of soft X-ray emission) signal could persist
steadily with $L\sim L_{\rm Edd}=10^{41}(M_{\rm h}/10^3M_\odot)$ erg/s
for at most tens of years; thereafter the flare would rapidly
fade. While the infall rate declines as $t^{-5/3}$, some material may
be stored for a longer time in an accretion disk. The IMBH luminosity
could then remain as high as $L_X\ge 10^{39}$ erg/s for several
hundreds of years after disruption.  In a given object, this
ultraluminous X-ray activity would have a duty cycle of order
$10^{-4}$.  Quiescent GCs, those with IMBHs now
starved of fuel, should greatly outnumber active ones; observational
constraints would not then be stringent until we had observed enough
candidates to constitute a proper ensemble average.
\end{abstract}
 

\section{Introduction}
Suggestive evidence has accumulated that intermediate mass black holes
exist in some globular clusters. There is some dynamical evidence for
a mass concentration within the central regions of some globular
clusters.  The dynamics of stars in the inner regions of nearby
clusters such as M15, G1 and $\omega$ Centauri suggest the presence of
black holes with masses of about $10^{3}$, $10^{4}$ and $4\times
10^4\;M_\odot$, respectively
\citep{ge2002,ge2003,geb2002,geb2005,po2006,noyola08}. The
peculiarities of an unresolved radio source in G1 indicate some unique
object at the center \citep{ul2007}. Recently, somewhat arguable
evidence has arisen for the presence of IMBHs in young star clusters,
where ultraluminous, compact X-ray sources (ULXs) have been
preferentially found to occur \citep{ze2002}. Their high luminosities
suggest that they are IMBHs rather than binaries containing a normal
stellar mass black hole \citep{po2004}. But before accepting this
conclusion (and dismissing alternative ideas) one would like some
independent corroboration of the IMBH hypothesis, or, conversely, some
way of ruling it out.

We are used to the idea that black holes are implicated in the most
powerful sources in the universe, and can (when accreting) be
ultra-efficient radiators. But there is, for example, no sign of such
activity in M15 - the X-ray upper limit is no more than $6 \times
10^{32}$ erg/s \citep{ho2003}. So could a black hole be so completely
starved of fuel that it does not reveal its presence? We do not
directly know how much gas there is near the IMBH \citep{p00,po2006},
and there is no a priori reason why this region should be {\it swept
  clean} of gas.  Gas that is lost from nearby stars \citep{ba2006} as
well as mass transfer from potentially bound companions
\citep{bl2006,pa2006,hopman04} can produce observable
signatures. However, it is uncertain whether gas can be adequately
supplied to explain ULX activity. The star density, on the other hand,
is much better known -- after all, if the stars were not closely
packed near the center of the cluster we would not have evidence for
the black hole at all.  As stellar orbits diffuse in phase space, it
therefore seems inevitable that some may wander sufficiently close to
the hole that they suffer tidal disruption.  When a star is disrupted,
there is bound to be some radiation from the sudden release of
gas. The flares resulting from a disrupted star could be the clearest
diagnostic of a IMBH's presence.

It is an intricate although tractable problem in stellar dynamics to
calculate the chance that a star passes within the IMBH's tidal radius
\citep{fr1976}. In a simple case when the velocities are isotropic,
the frequency with which a star would enter the zone of vulnerability
is $\xi \sim 10^{-7} M_{\rm h,3}^{4/3}(n_*/10^{6}\;{\rm
  pc^{-3}})(\sigma/ {\rm 10 \;km\; s^{-1}})^{-1}\;{\rm yr^{-1}}$,
where $n_*$ is the stellar number density in the cluster nucleus and
$M_{\rm h,3}$ is the black hole's mass in units of $10^3
M_\odot$. This estimate although simplified, agrees well with the
fiducial rates derived from detailed $N-$body simulations of
multi-mass star clusters containing IMBHs \citep{ba2004}.  It may
seem, however, that even the modest rate of stellar disruptions given
above could have conspicuous consequences. The debris from a disrupted
one solar-mass star per hundred million years, swallowed steadily with
10 per cent radiative efficiency, would yield a luminosity of $L_{\rm
  steady} \sim 6\times 10^{37}$ erg/s -- higher than is observed for
X-ray binaries in quiescence.  But in reality, one expects brighter
flares with short duty cycles, as the time it takes to digest or expel
the debris from one star is much shorter than the mean interval
between one stellar disruption and the next.  This {\it Letter} is
concerned with the observational manifestations of such phenomena,
with particular reference to globular clusters where the masses of the
black holes (if indeed present) are perhaps of order
$10^3-10^4M_\odot$.

\section{The tidal Disruption of solar-type star by an IMBH}
A star interacting with a massive black hole cannot be treated as a
point mass if it gets so close to the hole that it becomes vulnerable
to tidal distortions.  Such effects become important when the
pericenter distance becomes as small as the tidal radius: $R_{\rm T}
\simeq 5 \times 10^{11} M_{\rm h,3}^{1/3} (R_{\ast}/R_\odot)
(M_{\ast}/M_\odot)^{-1/3}\;{\rm cm}$. The gravitational radius,
$R_{\rm g} \simeq 1.5\times 10^{8} M_{\rm h,3}\;{\rm cm}$, scales with
mass, whereas the tidal radius goes only as the cube root. The tidal
forces at their ``surfaces'' are thus more gentle for black holes of
larger mass, and solar-type stars would be disrupted only after
passing irreversibly inside an ultra-massive hole's horizon: $M_{\rm
  h}\leq 7 \times 10^7\;M_\odot$.

When a rapidly changing tidal force starts to compete with a star's
self-gravity, the material of the star responds in a complicated way,
being stretched along the orbital direction and squeezed at right
angles to the orbit \citep{cl1983,re1988,ev1989}. To study this
problem, we use a three dimensional smoothed particle hydrodynamics
method (SPH) to solve the equations of hydrodynamics. Due to its
Lagrangian nature SPH is perfectly suited to follow tidal disruption
processes during which the corresponding geometry, densities and time
scales are changing violently. The SPH-formulation that we use in this
study is described in \citet{ro2008,rrh2008}. Common to all runs is
the initiation of the calculations with the star being place safely
outside $R_{\rm T}$ and set it onto a parabolic orbit so that $R_{\rm
  min}= R_{\rm T}/3$ for a $10^{3}M_\odot$ black hole. We have
considered three different initial conditions for the approaching
star, constructed here by solving the spherically symmetric Lane-Emden
equations: A [1, 1, 0.6] $M_\odot$ solar-type star modeled with a
polytropic equation of state with adiabatic index $\Gamma=[5/3, 1.4,
  1.4]$ and $R_\ast=[1,1,0.75] R_\odot$.

\begin{figure}
\plotone{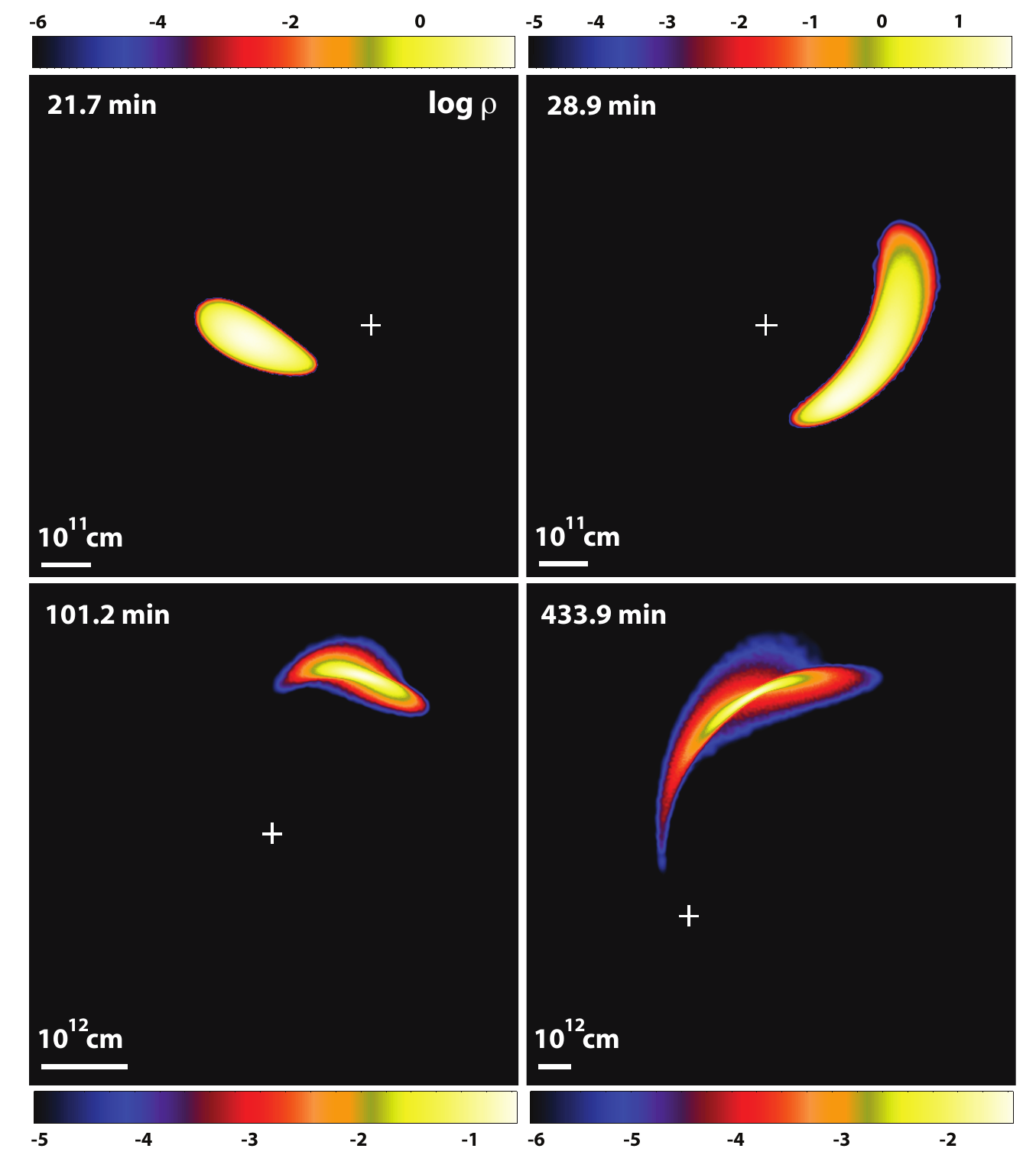}
\caption{A 1 $M_\odot$ solar-type star (modeled with more than $10^6$
  SPH particles) approaching a $10^3\;M_\odot$ black hole on a
  parabolic orbit with pericenter distance $R_{\rm min} = R_{\rm T}/3$
  is distorted, spun up during infall and then tidally disrupted. The
  panels show density cuts (in cgs units) through the orbital (xy-)
  plane before and after passage through pericenter.}
\label{fig1}
\end{figure}

The gross quantitative behavior of a solar-type star plunging deeply
within the tidal radius is examined here (the reader is refer to \S
\ref{ingesting} for a discussion on the role of the stellar density
structure in shaping the history of the mass accretion rate). Several
snapshots taken from our numerical simulations of a 1 $M_\odot$
solar-type star (modeled with $\Gamma=5/3$) are shown in
Figure~\ref{fig1}.  The tidal bulge raised on the star by the black
hole becomes of an order unity distortion near pericenter. The
resultant gravitational torque spins it up to a good fraction of its
corotation angular velocity by the time it gets disrupted.  This takes
place on a timescale comparable to the crossing time of the star
through periastron, $\Delta t \sim R_{\ast}/v_{\rm p} \sim 349.6
(M_{\ast}/M_\odot)^{-1/6}(R_{\ast}/R_\odot)^{3/2} M_{\rm
  h,3}^{-1/3}\;{\rm s}$.

\begin{figure}
\plotone{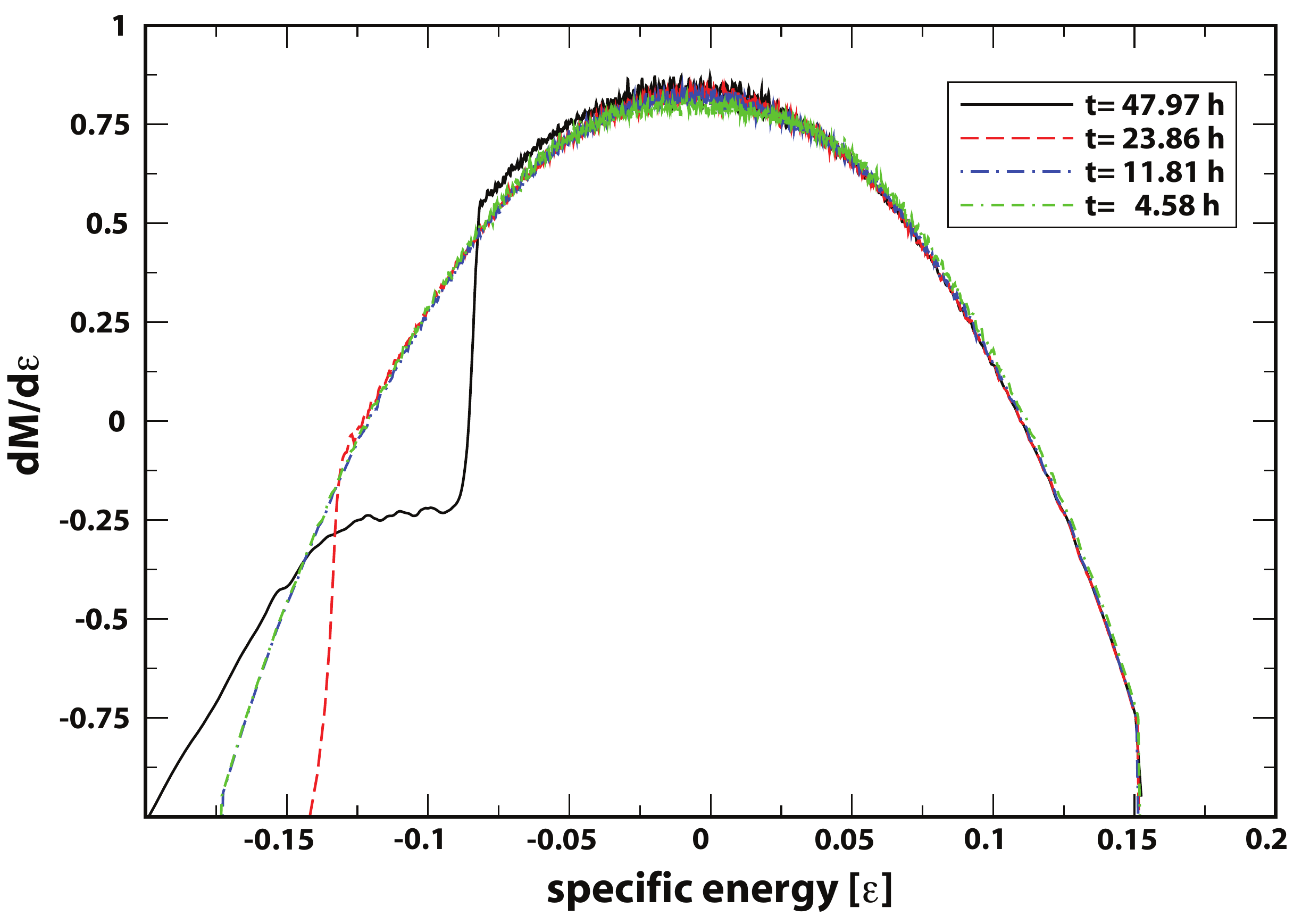}
\caption{Differential mass distributions in specific energy for the 1
  $M_\odot$ stellar debris. At $t=$47.97 h, the amount of material
  unbound is $\sim$ 47\% of the initial mass of the star.}
\label{fig2}
\end{figure}

The energy required to tear the star apart (that is the star's self
binding energy) is of order $M_\ast v_\ast^2$, where $v_\ast=(G
M_\ast/R_\ast)^{1/2}$. During tidal disruption, this energy is
supplied at the expense of the orbital kinetic energy, which at
pericenter $\sim R_{\rm T}$ is larger by $\sim (M_{\rm
  h}/M_\ast)^{2/3}$ \citep{re1988}. Figure~\ref{fig2} displays the
evolution of the differential mass distribution in specific energy for
the debris.  Disruption reduces the orbital energy by the binding
energy of the star $\epsilon_\ast \sim (G M_\ast/R_\ast) \approx
10^{-5}c^2$, which is much smaller than the specific kinetic energy at
pericenter. The variation of the specific energy in the released gas
is determined mainly by the relative depth of a mass element across
the disrupted star in the potential well of the black hole.  The
spread in this specific energy is of order $v\Delta v \sim 10^{-4}
M_{\rm h,3}^{1/3}c^2$, where $v\approx v_{\rm p}= c[R_{\rm g}/R_{\rm
    T}]^{1/2}$ and $\Delta v \approx v_\ast = [G
  M_\ast/R_\ast]^{1/2}$. This is much larger than $\epsilon_\ast$ and,
as a result, almost half the debris escapes on hyperbolic orbits with
speeds $\sim 3000\;M_{\rm h,3}^{1/6}\,{\rm km\,s^{-1}}$; the kinetic
energy output being $\sim 10^{50} M_{\rm h,3}^{1/3}\,{\rm erg}$
(comparable to the energy of a supernova). The material would be
concentrated in a {\it fan} close to the orbital plane. Adiabatic
cooling, as the material expands, severely reduces the internal
radiative energy content before the debris became optically thin.  The
escaping radiation from the unbound debris would therefore release
much less than the initial energy content of the star. There would
therefore be no conspicuous flare, until the bound debris fell back
onto the IMBH.

\section{Ingesting the Stellar Debris}\label{ingesting}
\begin{figure}
\plotone{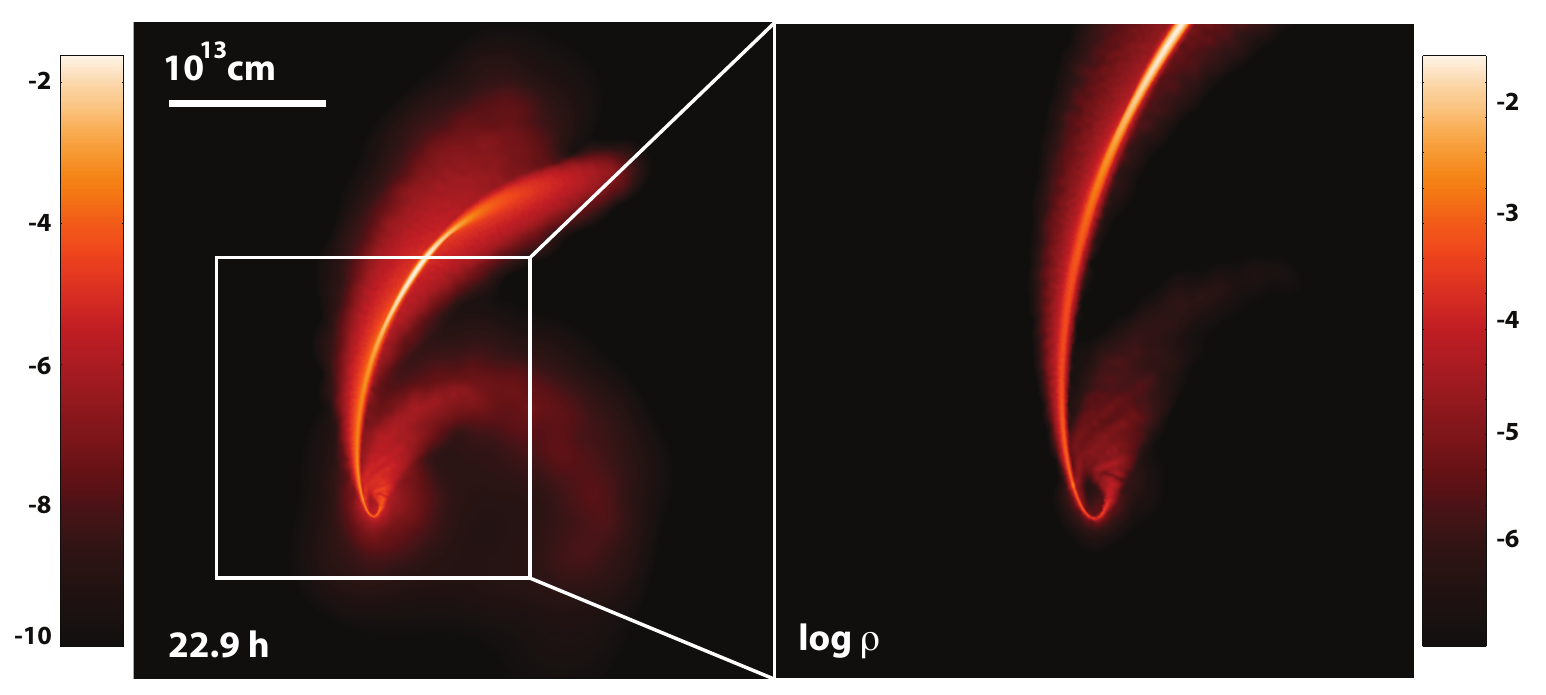}
\caption{Density cuts (in cgs units) in the orbital plane of the bound
  debris 22.9 hours after disruption. The most tightly bound debris
  would transverse an elliptical orbit with major axis $\sim 10^4
  R_{\rm g}$ before returning to $R\approx R_{\rm p}$, where radial
  focusing of orbits acts as an effective nozzle. These orbits are
  focused back into the original orbital plane at pericenter. This
  causes the formation of a pancake shock that weakly redistributes
  the orbital parameters and damps out some of the vertical motion. }
\label{fig3}
\end{figure}

The returning gas does not immediately produce a flare of activity
from the black hole. First material must enter quasi-circular orbits
and form an accretion torus \citep{ev1989}. Only then will viscous
effects release enough binding energy to power a flare. The bound
orbits are very eccentric, and the range of orbital periods is
large. The orbital semi-major axis of the most tightly bound debris is
$a\sim 10^4 M_{\rm h,3}^{-1/3} (R_\ast/R_\odot)
(M_\ast/M_\odot)^{-2/3}\; R_{\rm g}$, and the period is only $t_{\rm
  a}\sim 6300 (a/ 1.5\times 10^4 R_{\rm g})^{3/2} M_{\rm h,3}^{-1/2}$
s. If the gaseous debris suffered no internal dissipation due to high
viscosity or shocks, it would, after one or two orbital periods, form
a highly elliptical disk with a big spread in apocentric distances
between the most and least bound orbits, but where at pericenter the
orbits are all squeezed in a range $\delta R/ R\sim R_\ast/R_\tau= 0.1
(M_\ast/M_{\rm h,3})^{1/3}$. As the stream approaches pericenter, the
radial focusing of the orbits therefore acts as an effective nozzle
(Figure~\ref{fig3}). After pericenter passage, the outflowing gas is
on orbits which collide with the infalling stream near the original
orbital plane at apocenter, giving rise to an angular momentum
redistributing shock (Figure~\ref{fig3}) much like those in
cataclysmic variable systems. The debris raining down would, after
little more than its free-fall time, settle into a disk.

\begin{figure}
\plotone{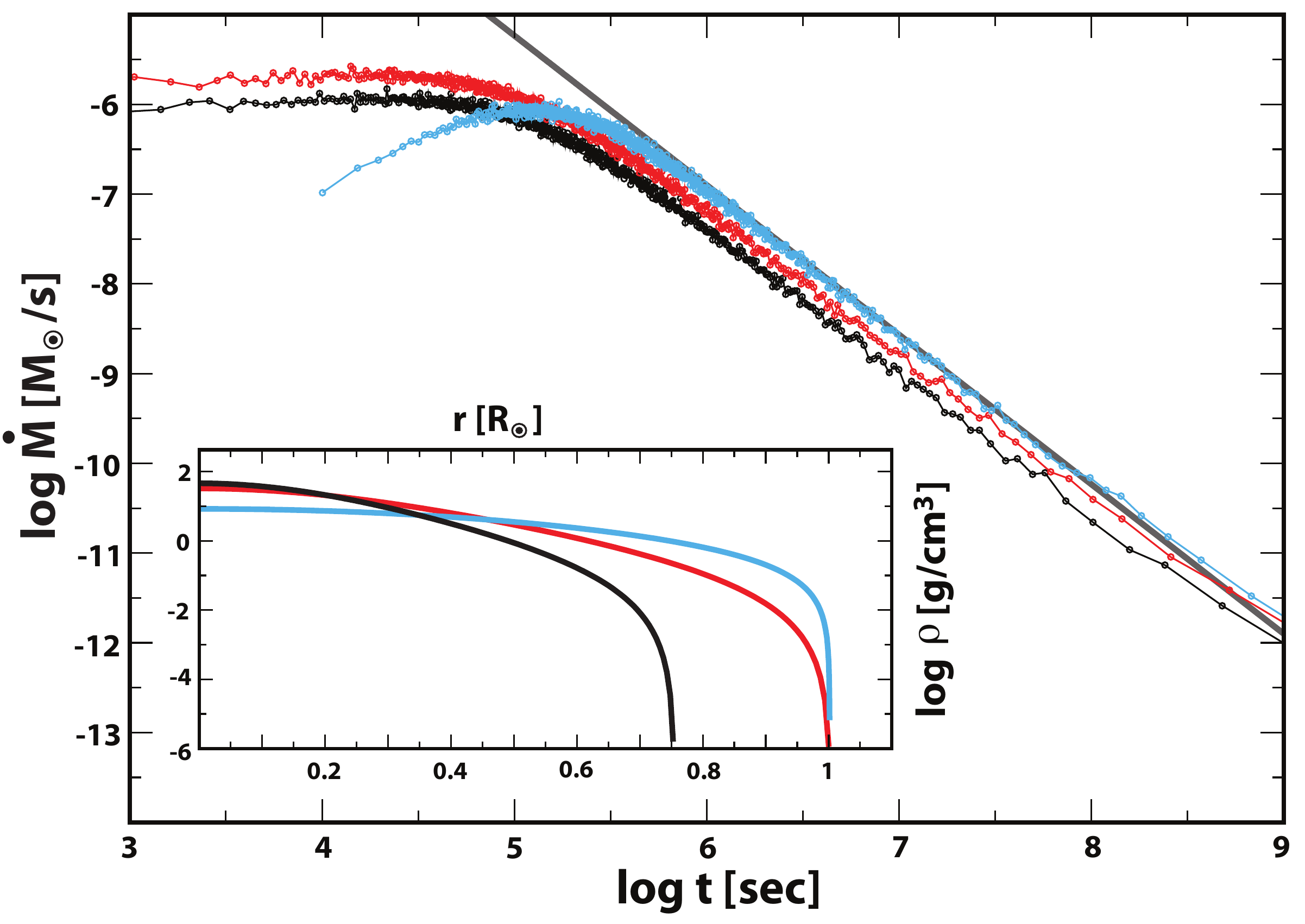}
\caption{The rates at which the stellar debris returns to the vicinity
  of the black hole for different types of stars: A [1,1,0.6]
  $M_\odot$ solar-type star modeled with $\Gamma=[5/3,1.4,1.4]$ shown
  as a solid [blue, red, black] line. The infalling matter at a rate
  that drops off roughly as $t^{-5/3}$ (gray line). {\it Inset}: the
  internal density structure of the various stars.}
\label{fig4}
\end{figure}

This orbiting debris starts forming a disk when the most highly bound
debris falls back.  The simulation shows that the first material
returns at a time $\leq t_{\rm a}$, with a peak infall rate roughly
given by $31 M_\odot\;{\rm yr^{-1}}$ (Figure~\ref{fig4}). Such high
infall rates are expected to persist, relative steadily, for at least
a few orbital periods, before all the highly bound material rains
down. Since the amount of highly bound material ($dM/d\epsilon$)
depends sensitively on the density structure of the star, we find that
more centrally condensed stars produce mass feeding rates that are
slightly larger in comparison to less centrally condensed stars of
comparable mass although they reach the self-similar $\propto
t^{-5/3}$ phase after fewer orbital periods.  The vicinity of the hole
would thereafter be fed solely by injection of the infalling matter at
a rate\footnote{Note that the rate at which mass returns to the black
  hole is only proportional to $t^{-5/3}$ when $dM/d\epsilon$ is
  constant \citep{re1988,phi1989}. Figure~\ref{fig2} displays the
  differential mass distributions in specific energy obtained by the
  simulation, which show that $dM/d\epsilon$ is nearly constant.} that
drops off roughly as $t^{-5/3}$ for $t \geq t_{\rm fb} \approx 10^5$
s. Once the torus is formed, it will evolve under the influence of
viscosity, radiative cooling winds and time dependent mass inflow.

 The black hole cannot accept matter, with a radiative efficiency
 $\epsilon=0.1\epsilon_{0.1}$, at a rate exceeding $\dot{M}_{\rm
   Edd}\approx2 \times 10^{-5} \epsilon_{0.1} M_{{\rm h},3}
 M_\odot\;{\rm yr^{-1}}$, without exceeding the black hole's Eddington
 luminosity. The rate at which the stellar debris returns to the
 vicinity of the black hole exceeds this limit: $\dot{M}\approx
 10^{6}\dot{M}_{\rm Edd}$ for $\epsilon=0.1$. Given that this high
 rate can only be sustained by infall of the highly bound material for
 a time $t_{\rm fb}$, we infer that, if the viscosity were high enough
 to process all the material within the infall timescale, then the
 luminosity of the hole could not remain as high as the Eddington
 luminosity for longer than $t_{\rm Edd}\approx 5 \times
 10^{3}\epsilon_{0.1}^{-3/5} M_{{\rm h},3}^{-3/5}\; t_{\rm fb}\sim 17
 \epsilon_{0.1}^{-3/5} M_{{\rm h},3}^{-3/5}\; {\rm yr}$. The viscosity
 would have to be implausibly low (i.e. the usual viscous dissipation
 time $t_{\rm d}$ for a thick disk would be $\alpha^{-1}$; $\alpha$
 would have to be below $10^{-5}$ for $t_{\rm d} \geq t_{\rm Edd}$)
 for the bulk of the mass to be stored for longer than $t_{\rm Edd}$
 in a reservoir at $R\sim R_{\rm T}$. A luminosity $\sim L_{\rm
   Edd}=10^{41}M_{{\rm h},3}\;{\rm erg\;s^{-1}}$ can therefore only be
 maintained for at most ten years; thereafter the flare would continue
 to fade as $t^{-3/5}$.  It is clear from the behavior of $\dot{M}$
 that most of the debris would be fed to the hole far more rapidly
 than it could be accepted if the radiative efficiency were high; much
 of the bound debris must either escape in a radiatively-driven
 outflow or be swallowed inefficiently.

\section{Discussion}
\subsection{Observability of  $L_{\rm Edd}$ Flares }
A distinctive consequence of a $10^3-10^4\;M_\odot$ IMBH's presence in
the centers of globular clusters would be the transient flares
produced as the bound debris from the disrupted solar-type stars is
swallowed, the luminosities being as high as $L_{\rm
  Edd}=10^{41}M_{\rm h,3}\;{\rm erg\;s^{-1}}$. The rise and the peak
bolometric luminosity can be predicted with some confidence. However,
the effective surface temperature (and thus the fraction of luminosity
that emerges predominantly in the soft X-ray band) is harder to
predict, as it depends on the size of the effective photosphere that
shrouds the hole. For $t\lesssim t_{\rm Edd}$, the effective surface
temperature should be $T \lesssim T_{\rm Edd} = (L_{\rm Edd}/4\pi
\sigma R_{\rm g}^2) ^{1/4}\sim 1 M_{\rm h.3}^{-1/4}$ keV. In a given
globular cluster, these flares would have a duty cycle of order $\xi
t_{\rm Edd} \sim 10^{-6} (\xi /10^{-7}\;{\rm yr^{-1}})(t_{\rm Edd}/10
{\rm yr})$, and as a result, quiescent GCs, those with IMBHs now
starved of fuel, should greatly outnumber active ones.

Given a globular cluster space density of $n_{\rm gc} \sim 4\;{\rm
  Mpc}^{-3}$ \citep{br2006}, we expect the density rate of flares with
$L=L_{\rm Edd}$ to be at most $\sim 4000 {\rm yr^{-1}}\;{\rm
  Gpc}^{-3}$, if $10^3-10^4\;M_\odot$ black holes were prevalent in
globular clusters. Therefore we would not yet expect to have detected
such a flare. Observational constraints on the presence of IMBH would
not then be stringent until a sufficiently large sample of galaxies
should be sampled in order to reveal some of their GC members in a
flaring state. Such objects should be searched for out to large
distances. However, assuming $L \sim 10^{41}$ erg s$^{-1}$ and $T \sim
T_{\rm Edd}$, we find that EXIST \citep{grindlay04} would only be able
to detect such a flaring event out to a distance of about 1 Mpc.

\subsection{ULX activity}
A further question is how fast the luminosity fades after the
flare. This is important because we want to know whether the IMBH has
faded below ULX levels before the next stellar disruption occurs. The
answer to this question depends on how long it takes the last residues
of the star to be ingested. One expects the infall rate $\dot{M}$ to
decline as $t^{-5/3}$ for late times. Some material may, however, be
stored for a longer time in an accretion disk. This is mainly because
the specific angular momentum for Keplerian orbits grows $\propto
r^{1/2}$, so angular momentum transport via disk viscosity requires
that 10\% of the debris goes out to $10^3\;R_{\rm T}$ and 1\% to
$10^5\;R_{\rm T}$ before being swallowed.

The evolution of the bound debris onto a black hole for $t\geq t_{\rm
  a}$ has been studied by \citet{ca1990} using a time dependent
$\alpha$-disk model. Cannizzo and collaborators concluded that the
black hole's luminosity fades more slowly: $L \propto t^{-1.2}$ and
depends only weakly on the disk viscosity. The IMBH luminosity could
then remain as high as $L_X\ge 10^{39}\;{\rm erg\;s^{-1}}$ for several
hundreds of years after disruption.  In a given object, this ULX
activity would have a duty cycle of order $10^{-4}$ unless the rate of
disruptions is much larger than $10^{-7}$ or the viscosity were so low
that the bulk of the mass could be stored for many thousands of years
at $R\sim R_{\rm T}$ (where the dynamical timescale is only a few
hours).

\subsection{Relevance to GCs and  IMBH Growth}
The mass fraction that is ejected rather than ingested, although less
spectacular than the accretion-powered flares could nonetheless have
an important influence on the energy balance within a GC - similar
than a supernova exploding in the same volume. When the star is
disrupted in a single flyby, about half the debris is ejected in a
fast moving spray of gas; the kinetic energy output being $\sim
10^{50} M_{\rm h,3}^{1/3}\,{\rm erg}$. The ejecta would be braked and
their kinetic energy thermalized, as they ran into the diffuse gas
(probably originating from stellar mass loss). If this material is
effectively shock heated, it could contribute to the X-ray and radio
luminosity of the GC long after the star has been disrupted. Stellar
disruption may have other implications -- for instance, to the
dynamics of line emitting regions within GCs \citep{ch2008}.

The most tightly bound debris, on the other hand, would traverse an
elliptical orbit before returning to $R_{\rm T}$.  There would
therefore be no conspicuous flare until, as discussed above, the bound
debris fall back into the hole. Electron scattering opacity almost
certainly dominates the radiative transfer and the photons will be
trapped out to a radius $R_\tau =\dot{M}\kappa_\tau/4\pi c$. Energy
will be dissipated at a supercritical rate as the material swirls
closer to the hole, some gas may then be ejected in a radiation-driven
wind. In principle it is possible for only a fraction $\epsilon^{-1}
(R_{\rm g}/R_{\rm T}) \sim 0.004
\epsilon_{0.1}(R_\ast/R_\odot)^{-1}(M_\ast/M_\odot)^{1/3}M_{\rm
  h,3}^{2/3}$ to be actually swallowed, all the remainder being
ejected. If a fraction $f$ of the mass from disrupted stars is
accreted onto the black hole, the average rate of tidal disruptions
required to form an IMBH out of a $\sim 50 M_\odot$ progenitor would
need to be $\ge 10^{-6} M_{\rm h,3} (f/0.1)^{-1} (\tau_{\rm
  GC}/10^{10}{\rm yr})^{-1} {\rm yr}^{-1}$, where $\tau_{\rm GC}$ is
the globular cluster age. In such cases, the density would, however,
be high enough that runaway merging of high-mass main-sequence stars
could lead directly to the formation of an IMBH \citep{ba2006,po2004}.

\acknowledgments We thank H. Baumgardt, L. Chomiuk, J. Strader, J.
Guillochon, P. Hut, E. Noyola, C. Hopman, J. Kalirai, D. Kasen and
M. Rees for useful discussions and the referee for constructive
suggestions. E. R. acknowledges support from the DOE (SciDAC;
DE-FC02-01ER41176) and NSF (0521566). The simulations presented in
this paper were performed on the JUMP computer of the
H\"ochstleistungsrechenzentrum J\"ulich.

\acknowledgments

\newpage

\newpage

\newpage

\newpage

\end{document}